# Multiloop Calculations in QED by Superparticle Path Integrals

Michael G. Schmidt

*Institut für Theoretische Physik*
*Universität Heidelberg*
*Philosophenweg 16*
*69120 Heidelberg*

Christian Schubert

*Institut für Hochenergiephysik Zeuthen*
*DESY Deutsches Elektronen-Synchrotron*
*Platanenallee 6*
*15738 Zeuthen*

**August 1994**

## Abstract

We use the worldline path integral approach to the Bern-Kosower formalism for extending this formalism to the calculation of higher order corrections to $N$ – photon scattering. The case of the two-loop QED $\beta$-function is considered in detail.

We report on some progress we have recently made concerning the generalization of the Bern-Kosower formalism beyond the one-loop level. This formalism, which offers an alternative to Feynman diagram calculations, was originally developed by analyzing the infinite string-tension limit of certain superstring amplitudes [1], and its usefulness for one-loop calculations in gauge theories has been demonstrated by the calculation of both five-point gluon [2] and four-point graviton [3] amplitudes.

Considering the difficulty of doing two-loop QCD calculations by standard methods, it would be obviously desirable to have a higher-order generalization of this method. While some progress has been made toward this goal in the original framework of Bern and Kosower [4], a derivation of two-loop rules for nonabelian gauge theory has, so far, remained elusive.

One may hope that something can be learned for this problem from the simpler case of photon scattering in QED, which we will consider in this paper.

For this purpose, we use the reformulation of the Bern-Kosower formalism due to Strassler [5]. This approach is based on the worldline path integral representation for the one-loop effective action induced by a spinor loop,

$$\Gamma[A] = -2\int_0^\infty \frac{dT}{T} e^{-m^2 T} \int \mathcal{D}x \mathcal{D}\psi \, \exp\Big[-\int_0^T d\tau \Big(\frac{1}{4}\dot{x}^2 + \frac{1}{2}\psi\dot{\psi} + igA_\mu \dot{x}^\mu - ig\psi^\mu F_{\mu\nu}\psi^\nu\Big)\Big], \tag{1}$$

where the $x^\mu(\tau)$'s are the periodic functions from the circle with circumference $T$ into $D$ – dimensional Euclidean spacetime, and the $\psi^\mu(\tau)$'s their antiperiodic supersymmetric partners. The case of scalar electrodynamics corresponds to deleting the fermionic path integral and the global factor of $-2$.

The "string-inspired" way of evaluating this type of path integral [5, 6] consists in extracting the integration over the center of mass $x_0$,

$$\int \mathcal{D}x = \int dx_0 \int \mathcal{D}y, \quad \int_0^T d\tau\, y^\mu(\tau) = 0, \tag{2}$$

and evaluating the path integrals over $y$ and $\psi$ by Wick contractions as in a one-dimensional field theory on the circle. The Green functions to be used are those adapted to the (anti-) periodicity conditions,

$$\begin{aligned}
\langle y^\mu(\tau_1) y^\nu(\tau_2) \rangle &= -g^{\mu\nu} G_B(\tau_1, \tau_2) = -g^{\mu\nu}\Big[|\tau_1 - \tau_2| - \frac{(\tau_1 - \tau_2)^2}{T}\Big], \\
\langle \psi^\mu(\tau_1) \psi^\nu(\tau_2) \rangle &= \frac{1}{2} g^{\mu\nu} G_F(\tau_1, \tau_2) = \frac{1}{2} g^{\mu\nu} \operatorname{sign}(\tau_1 - \tau_2).
\end{aligned} \tag{3}$$

As shown in [5], explicit execution of the Wick contractions is necessary only for the bosonic path integral; once this has been done, the fermionic contributions may be incorporated using a simple substitution rule derived from worldline supersymmetry.

One-loop scattering amplitudes are obtained by specializing the background to a finite sum of plane waves. In the case of QED, this procedure then leads to the same integral representations for one-loop photon scattering amplitudes as application of the Bern-Kosower rules.

To extend this calculus to higher orders, one needs to know the generalizations of the above Green functions on the circle to higher order graphs. The simplest case to consider here – which is also the one relevant to photon scattering – is the case of a loop with insertions. In [7] it had been shown that the insertion of a scalar propagator into a loop between points $x(\tau_a)$ and $x(\tau_b)$, written in the Schwinger proper-time representation

$$\int_0^\infty d\bar{T} e^{-m^2 \bar{T}} (4\pi\bar{T})^{-\frac{D}{2}} \exp\left[-\frac{(x(\tau_a) - x(\tau_b))^2}{4\bar{T}}\right], \tag{4}$$

changes the Green function for the loop path integral into

$$G_B^{(1)}(\tau_1, \tau_2) = G_B(\tau_1, \tau_2) + \frac{1}{2} \frac{[G_B(\tau_1, \tau_a) - G_B(\tau_1, \tau_b)][G_B(\tau_2, \tau_a) - G_B(\tau_2, \tau_b)]}{\bar{T} + G_B(\tau_a, \tau_b)} \tag{5}$$

A similar formula exists for multiloop insertions.

For extension to scalar electrodynamics, we have to replace the inserted scalar propagators by photon propagators. As is well-known from Wilson-loop theory, insertion of a photon propagator (in Feynman gauge) can be written as

$$\frac{g^2}{8\pi^2} \int_0^T d\tau_a \int_0^T d\tau_b \frac{\dot{x}^\mu(\tau_a)\dot{x}_\mu(\tau_b)}{(x(\tau_a) - x(\tau_b))^2} \quad . \tag{6}$$

The transition to spinor electrodynamics may then be simply achieved by supersymmetrization, as will be explained in more detail elsewhere [8].

As an illustration, let us recalculate the two-loop $\beta$ – function for spinor QED. In a Feynman diagram calculation of this quantity, one would have to calculate the three diagrams of fig. 1 using some regularization, say dimensional regularization, and then extract their $\frac{1}{\epsilon}$ – poles. In the present formalism, those diagrams are combined into one integral, which has the useful consequence that we will not need a regulator except for the overall divergence.

If one is only interested in the $\beta$ – function contribution, things may be simplified by choosing a constant background field $F_{\mu\nu}$, and calculating the coefficient of the induced $F_{\mu\nu}F^{\mu\nu}$ – term. One can then choose a gauge such that $A_\nu = \frac{1}{2}x^\mu F_{\mu\nu}$, which makes the bosonic interaction term equal to $i\frac{g}{2}\dot{x}^\mu F_{\mu\nu} x^\nu$.

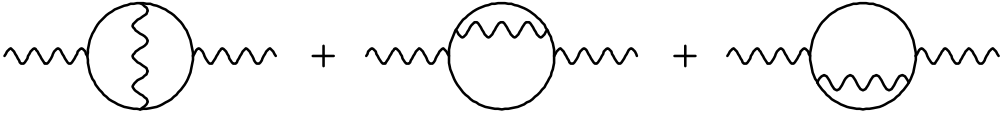

Figure 1: Diagrams contributing to the two-loop vacuum polarization

We first calculate the purely bosonic contributions, which are, as for scalar electrodynamics, obtained by inserting into the one-loop path integral the worldline current-current interaction term eq.( 6). The denominator is written in the proper-time representation, and absorbed into the worldline Green function as in the scalar case; the numerator remains, and participates in the Wick contractions. Expansion of the interaction exponential to second order yields

$$\Gamma_{bos}[F] = -2(4\pi)^{-4} \int dx_0 \int_0^\infty \frac{dT}{T^3} e^{-m^2 T} \int_0^\infty d\bar{T} \int_0^T d\tau_a \int_0^T d\tau_b [\bar{T} + G_{Bab}]^{-2}$$
$$\times (-\frac{g^2}{2})\frac{g^2}{2} \int_0^T d\tau_1 \int_0^T d\tau_2 \frac{1}{4} \langle \dot{y}_1^\mu F_{\mu\nu} y_1^\nu \dot{y}_2^\alpha F_{\alpha\beta} y_2^\beta \dot{y}_a^\lambda \dot{y}_{b\lambda} \rangle \qquad (7)$$

(note the appearence of the two-loop determinant factor $[\bar{T} + G_B(\tau_a,\tau_b)]^{-2}$ [7]). The Wick contraction of $\langle \dot{y}_1^\mu y_1^\nu \dot{y}_2^\alpha y_2^\beta \dot{y}_a^\lambda \dot{y}_{b\lambda} \rangle$ is then performed, using the two-loop Green function eq.( 5), and the result written out in terms of the bosonic one-loop Green function and its derivatives.

To include the fermionic contributions, one now eliminates all factors of $\ddot{G}_B$ by partial integrations with respect to $\tau_1, \tau_2, \tau_a, \tau_b$, and uses the one-loop substitution rule, replacing, for example,

$$\dot{G}_{B12} \dot{G}_{B21} \dot{G}_{Bab} \dot{G}_{Bba} \to (\dot{G}_{B12}\dot{G}_{B21} - G_{F12}G_{F21})(\dot{G}_{Bab}\dot{G}_{Bba} - G_{Fab}G_{Fba}) \qquad (8)$$

etc. A sixfold integral remains to be done (fig. 2), the integrand being a polynomial in the various $G_{Bij}, \dot{G}_{Bij}, G_{Fij}$, and in $[\bar{T} + G_B(\tau_a,\tau_b)]^{-1}$.
In particular, the integrations over $\tau_1, \tau_2$ are polynomial, and the $\bar{T}$ – integration may be made so by a change of variables

$$\bar{T} \to \beta(\bar{T}) := \frac{G_B(\tau_a,\tau_b)}{\bar{T} + G_B(\tau_a,\tau_b)} \qquad . \qquad (9)$$

The result of these three integrations is a number of terms proportional to $G_B(\tau_a,\tau_b)^{-2}$, $G_B(\tau_a,\tau_b)^{-1}$ and $G_B(\tau_a,\tau_b)^0$. They all cancel, however, except the constant ones, rendering the $\tau_a, \tau_b$ – integrals trivial. To extract the divergence of the remaining proper-time integral, one may use a Pauli-Villars regulator $\Lambda$, yielding

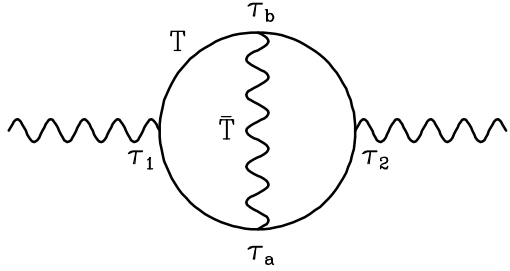

Figure 2: Definition of the six integration parameters

$$\int_0^\infty \frac{dT}{T} e^{-m^2 T} \sim \ln \frac{\Lambda^2}{m^2} \qquad (10)$$

The final result for the divergent part of the two-loop effective action becomes

$$\Gamma^{(2)} \sim -\ln \frac{\Lambda^2}{m^2} (4\pi)^{-4} g^4 \int dx_0 F_{\mu\nu} F^{\mu\nu} \quad , \qquad (11)$$

from which one can derive the two-loop $\beta$ – function coefficient in the usual way, and with the usual result

$$\beta^{(2)}(\alpha) = \frac{\alpha^3}{2\pi^2} \quad . \qquad (12)$$

Let us summarize the peculiarities of this calculation:

i) All integrations have been polynomial.

ii) The three diagrams of fig. 1 have been combined into one calculation.

iii) As a consequence, no regularization has been necessary except for the final proper-time integral.

While this calculation cannot yet be considered a serious test of the formalism, those properties are certainly encouraging.